\documentstyle[prc,aps]{revtex}
\begin{document}
\draft
\title{CORRELATED $\bbox{\pi \rho}$ EXCHANGE IN THE $\bbox{NN}$
INTERACTION}
\author{G. Janssen, K. Holinde, and J. Speth}
\address{Institut f\"{u}r Kernphysik, Forschungszentrum J\"{u}lich GmbH,
         \\ D-52425 J\"{u}lich, Germany}
\maketitle
\begin{abstract}
We evaluate the contribution to the nucleon-nucleon interaction due to
correlated $\pi\rho$ exchange in the $\pi$, $\omega$, and $A_1$/$H_1$
channels by means of dispersion-theoretic methods based on a realistic
meson exchange model for the interaction between $\pi$ and $\rho$
mesons. These processes have substantial effects: In the pionic channel
it counterbalances the suppression generated by a soft $\pi NN$ form
factor of monopole type with a cutoff mass of about 1 GeV; in the
$\omega$-channel it provides nearly half of the empirical repulsion,
leaving little room for explicit quark-gluon effects.
\end{abstract}

\section {Introduction}

The study of meson-meson systems and their role in low and medium energy
physics is of twofold interest. First, from a more basic viewpoint, the
investigation of meson-meson interactions provides important information
about the fundamental structure of strong interactions.  This is
especially true for the lightest system consisting of two
pions. $\pi\pi$ scattering at low energies is determined by chiral
symmetry and therefore plays a dominant role in chiral perturbation
theory. For higher energies the study of $\pi\pi$ as well as $\pi\eta$
scattering (including the coupling to the $K\overline K$ channel)
provides essential information on the nature of scalar
resonances. Within a meson exchange $\pi\pi\, (\pi\eta)/K\overline K$
model the $f_0(980)$ turns out to be a $K\overline K$ bound state and
the $a_0(980)$ a $K\overline K$ threshold effect \cite{Janssen95}.
Second, the inclusion of meson-meson correlations is mandatory for
numerous hadronic processes.  In models of nucleons consisting of a
quark-gluon `core' and a meson cloud, meson-meson couplings have a large
impact on the structure of nucleon form factors.  In meson-exchange
models of the $NN$ interaction, which essentially include pseudoscalar
and vector mesons, one has for consistency not only to include the
meson-nucleon, but also the meson-meson interaction. Indeed, the
correlated $2\pi$-exchange contribution is of outstanding importance,
providing the main part of the intermediate-range attraction.

In this work we want to demonstrate the important role of the
contribution to the nucleon-nucleon \cite{Machleidt87} interaction
provided by the exchange of a correlated $\pi\rho$ pair (A corresponding
letter has already been published~\cite{JanssenPRL94}). The starting
points are open questions concerning the structure of the $\pi NN$
vertex. Besides the coupling constant, the $\pi NN$ vertex function
(like all baryon-baryon meson vertices) contains a form factor
parametrizing the extended hadron structure and characterized, in a
monopole parametrization, by a cutoff mass $\Lambda_{\pi NN}$. Within
the (full) Bonn potential, the best fit to the $NN$ data requires a
value of $\Lambda_{\pi NN}$ = 1.3 GeV; the resulting form factor
modifies the one-pion-exchange potential for small distances (r$\le$ 1
fm) only. This ensures a tensor force which is strong enough to
reproduce the deuteron properties, especially the D/S ratio and the
quadrupole moment \cite{Ericson83,Ericson85}.  However such a hard form
factor is in contradiction to information from other sources
\cite{Liu94,Coon81,Coon90,Thomas89}, which all favor a considerably
lower value of $\Lambda_{\pi NN}\simeq$0.8 GeV, i.e.\ a rather `soft'
form factor.  Obviously the solution of this problem requires the
inclusion of additional (short ranged) tensor contributions in the Bonn
potential, which compensate for the effect of such a soft $\pi NN$ form
factor.  The exchange of a correlated $\pi\rho$ pair is a natural
candidate for such a contribution.  Indeed, the inclusion of
uncorrelated $\pi\rho$ processes in the full Bonn potential already led
to a reduction of the $\pi NN$ cutoff mass from 1.75 GeV (in the
framework of a simple one-boson-exchange model
(OBEPT)~\cite{Machleidt87} defined likewise in time-ordered perturbation
theory) to 1.3 GeV. Thus a further reduction of this value is to be
expected if correlated $\pi\rho$ exchange is included.  This is anyway
required within the strategy advocated in the full Bonn potential:
namely to group $\pi\pi$ and corresponding $\pi\rho$ contributions
together because of their counterstructure in the tensor channel.
However, while the Bonn potential contains already correlated $\pi\pi$
exchange (in terms of sharp mass $\sigma'$ exchange), the corresponding
$\pi\rho$ process is not included so far.

The evaluation of correlated $\pi\rho$ exchange (Fig.~\ref{figintI})
requires the knowledge of the interaction between $\pi$ and $\rho$
mesons. We have recently derived a corresponding meson-theoretical model
\cite{JanssenPRC94} for $\pi\rho$ scattering which provides good
agreement with the existing empirical information. It is, however, not
completely crossing symmetric, so that a direct evaluation of
Fig.~\ref{figintI}(d) based on this $\pi\rho$ interaction is not
possible. Therefore, as in the $\pi\pi$ case \cite{Kim94}, we first
evaluate the amplitude $N\overline N \to \pi\rho \to N\overline N$
including $\pi\rho$ correlations. In a second step we use
dispersion-theoretic methods to transform this amplitude into the $s$
($NN$) channel and in this way obtain the contribution of
Fig.~\ref{figintI}(d).

A main result will be that the exchange of a correlated $\pi\rho$ pair
indeed generates a contribution to the $NN$ potential of sizable
strength. In the pionic channel it leads to a tensor force component
which is strong enough to cancel the effect of a soft form factor with
$\Lambda_{\pi NN}\simeq$1 GeV.  Moreover, it provides nearly half of the
empirical repulsion in the $\omega$-channel.

The structure of the paper is as follows: Sect.~\ref{sect:form} provides
the basic formalism, together with our model for the $N\overline N \to
\pi\rho$ amplitude; Sect.~\ref{sect:results} presents and discusses the
results, in the various ($\pi$, $\omega$, $A_1/H_1$) channels
considered; finally, Sect.~\ref{sect:concl} contains some concluding
remarks.

\section {Formalism}
\label{sect:form}

In the following we outline the formalism which is used to evaluate the
correlated $\pi\rho$ exchange contribution to the $NN$ interaction.

\subsection{Dispersion relation for the
$NN\to NN$ amplitude}

For $NN\to NN$ and $N\overline N \to N\overline N$ scattering the field
theoretical scattering amplitude $T$ is related to the standard
$S$-matrix by

\begin{equation}
S_{fi}=\delta_{fi}-i(2\pi)^{-2}\delta^{(4)}(p_1'+p_2'-p_1-p_2) \left (
\frac{m_N^4}{E_1'E_2'E_1E_2} \right) ^{\frac{1}{2}} T_{fi} \ .
\end{equation}

If we neglect isospin for the moment, the $s$-channel ($NN\to NN$)
amplitude can be written as

\begin{equation}
T_s(p_1',p_2';p_1,p_2)=\overline u(p_1',\lambda_1') \overline
u(p_2',\lambda_2') \hat T u(p_1,\lambda_1)u(p_2,\lambda_2) \quad,
\label{invampI}
\end{equation}
\noindent
where

\begin{equation}
u(p,\lambda)=\sqrt{{E(p)+m_N\over 2m_N}}\left (
\begin{array}{c}
1 \\{2\lambda p \over E(p)+m_N}\end{array}\right ) |\lambda>
\end{equation}
is a Dirac helicity spinor normalized to $\overline u u =1$. (The spin
 dependence is suppressed on the left hand side of Eq.~(\ref{invampI}).)
 For on-shell scattering, $\hat T$ can be expressed as a linear
 combination of five invariant operators $\hat C_j$; the expansion
 coefficients $c_j$ are scalar functions of the Mandelstam variables
 $s\equiv (p_1+p_2)^2$ and $t\equiv (p_1'-p_1)^2$.  ($u$ is not
 independent, but given by $u=4m_N^2-s-t$.)  $\hat T$ can then be
 written as

\begin{equation}
\hat T=\sum^5_{j=1} c_j(t,s)\hat C_j \ .
\label{expans}
\end{equation}

In contrast to our former work dealing with correlated $\pi\pi$ exchange
\cite{Kim94} we now use, instead of the so-called perturbative
invariants (see Ref.~\cite{Kim94}) the Fermi-invariants

\begin{eqnarray}
S &=& (I)^{(1)}\, (I)^{(2)} \nonumber \\ P &=& (\gamma_5)^{(1)}\,
(\gamma_5)^{(2)} \nonumber \\ V &=& (\gamma^\mu)^{(1)}\,
(\gamma_\mu)^{(2)} \nonumber \\ A &=& (\gamma_5\gamma^\mu)^{(1)}\,
(\gamma_5\gamma_\mu)^{(2)} \nonumber \\ T &=& (\sigma^{\mu\nu})^{(1)}\,
(\sigma_{\mu\nu})^{(2)} \ \ .
\label{Fermi}
\end{eqnarray}

\noindent
where $\sigma_{\mu\nu}\equiv\frac{i}{2}[\gamma_\mu,\gamma_\nu]$.

Correspondingly, the amplitude for the $t$-channel ($N\overline N \to N
\overline N$) process defined in Fig.~\ref{figone} then reads

\begin{equation}
T_t(-p_1',p_2';p_1,-p_2)=\overline v(-p_1',\bar\lambda_1') \overline
u(p_2',\lambda_2') \hat T v(p_1,\lambda_1)u(-p_2,\bar \lambda_2) \ .
\label{invampII}
\end{equation}
Here
\begin{equation}
 v(- p,\lambda)=\sqrt{{E(p)+m_N\over 2m_N}} \left (
\begin{array}{c}
{p\over E(p)+m_N} \\ -2\lambda\end{array}\right ) |-\lambda>
\end{equation}
is the Dirac spinor for an antiparticle. Due to crossing symmetry $\hat
T$ can be represented in the same way as before (Eq.~(\ref{expans})),
with precisely the same functions $c_j$, however in a different $s$, $t$
domain obtained by replacing $p_1'$ by $-p_1'$ and $p_2$ by $-p_2$.

The functions $c_j$ arising from (correlated) $\pi\rho$ exchange are
assumed to fulfill a dispersion relation over the unitarity cut,

\begin{equation}
c_j(t,s)=\frac{1}{\pi}\int^\infty_{(m_\pi+m_\rho)^2} \frac{ {\rm Im}\,
c_j(t',s)} {t'-t-i\epsilon}dt' \ .
\label{Disp}
\end{equation}

\noindent
(Throughout we take the $\rho$ to be a stable particle with $m_\rho$=769
MeV.)  Thus the $c_j$ can be determined if their imaginary part is known
in the pseudophysical region ($t'\ge (m_\pi+m_\rho)^2$) of the
$t$-channel reaction and for $s\ge 4m_N^2$.

\subsection{Determination of the spectral functions from unitarity}

The required information about the spectral functions
$\rho_j(t',s)\equiv {\rm Im}\ c_j(t',s)$ can be obtained from the
relevant unitarity relation (cf.\ Fig.~\ref{figtwo})

\begin{equation}
i<N\overline N |\hat T-\hat T^\dagger|N \overline N> =\sum_{\pi\rho}
\Omega_{\pi\rho}<N\overline N |t^\dagger|\pi\rho> <\pi\rho |t|N
\overline N> \delta^{(4)}(k_1+k_2+p_1'-p_1)
\label{unit}
\end{equation}

\noindent
where $\Omega_{\pi\rho}$ is a $\pi\rho$ phase-space factor.  We first do
a partial wave decomposition,

\begin{eqnarray}
T_t ({\bf p'} \lambda_N'\lambda_{\overline N}'; {\bf p}
\lambda_N\lambda_{\overline N};\sqrt t) &=&\frac{1}{4\pi}\sum_J\,(2J+1)
d^J_{\lambda\lambda'}(\cos\vartheta) T_t^J ( p'
\lambda_N'\lambda_{\overline N}'; p \lambda_N\lambda_{\overline N};\sqrt
t)\nonumber \\ t({\bf k} \lambda_\rho; {\bf p}
\lambda_N\lambda_{\overline N};\sqrt t) &=&\frac{1}{4\pi}\sum_J\,(2J+1)
d^J_{\lambda\bar\lambda}(\cos\bar\vartheta) t^J(k \lambda_\rho; p
\lambda_N\lambda_{\overline N};\sqrt t)
\label{PWZ}
\end{eqnarray}
with
\begin{equation}
 \lambda'\equiv \lambda_N'-\lambda_{\overline N}',\quad
\bar\lambda\equiv \lambda_\rho-\lambda_\pi=\lambda_\rho, \quad
\lambda\equiv \lambda_N-\lambda_{\overline N}\quad,
\end{equation}
${\bf p}$, ${\bf p'}$, ${\bf k}$ being the relative 3-momenta in the
 center-of-mass (cm) system and the angles $\vartheta=\angle({\bf
 p},{\bf p'})$, $\bar\vartheta=\angle({\bf p},{\bf k})$.  After
 transformation into LSJ basis Eq.~(\ref{unit}) goes into

\begin{eqnarray}
&&{\rm Im}[T_t^J (p_0,L'\,S';p_0,L\,S;\sqrt t)] \nonumber \\ &&= -
C\sum_{L_{\pi\rho}}[t^J(k_0,L_{\pi\rho}\,1;p_0,L'\,S';\sqrt t)]
^\dagger\ t^J(k_0,L_{\pi\rho}\,1;p_0,L\,S;\sqrt t) \equiv\
^J\!N(L'S';LS)
\label{UniRel}
\end{eqnarray}

\noindent
where $C\equiv k_0/(32\pi^2\sqrt t)$ and $p_0$, $k_0$ denote the
on-shell momenta of the $N\overline N$ and $\pi\rho$ system,
respectively.

As for the $2\pi$-exchange case, we want to restrict ourselves to the
$J=0,1$ $\pi\rho$ exchange contributions, which act in channels
corresponding to the quantum numbers of the pion ($J=0$), and the
$\omega$, $A_1$, and $H_1$ meson ($J=1$). Table~\ref{Tabone} shows the
quantum numbers and possible transitions from the $N\overline N$ to the
$\pi\rho$ system obtained from the conditions

\begin{eqnarray}
(-1)^{L_{N\overline N}+1}\ \ \ \ \ \ &=&P_\pi P_\rho(-1)^{L_{\pi\rho}}
=(-1)^{L_{\pi\rho}}\\ (-1)^{L_{N\overline N}+S_{N\overline N}+I}&=&G_\pi
G_\rho =-1
\end{eqnarray}

\noindent
due to parity and G-parity conservation.  We therefore obtain for the
different relevant channels

\begin{equation}
\begin{array}{l}
^0N(00;00) = -C (t^0_{0+})^\dagger t^0_{0+}
\hspace{2.6cm} \pi \\ \\ \left.
\begin{array}{l}
^1N(01;01) = -C (t^1_{-1})^\dagger t^1_{-1}
\hspace{1cm} \\ ^1N(21;01) = -C (t^1_{+1})^\dagger t^1_{-1}
\hspace{1cm} \\ ^1N(01;21) = -C (t^1_{-1})^\dagger t^1_{+1}
\hspace{1cm} \\ ^1N(21;21) = -C (t^1_{+1})^\dagger t^1_{+1}
\hspace{2cm} \\
\end{array}
\right \} \omega \\ \\ ^1N(11;11) = -C [(t^1_{1-})^\dagger
t^1_{1-}+(t^1_{1+})^\dagger t^1_{1+}] \ \ A_1 \\ ^1N(10;10) = -C
[(t^1_{0-})^\dagger t^1_{0-}+(t^1_{0+})^\dagger t^1_{0+}] \ \ H_1 \\
\end{array}
\label{LSJamps}
\end{equation}

The knowledge of $^JN$ determines the spectral functions. With the help
of Appendix A, we have explicitly

\begin{eqnarray}
&&\rho^\pi_P(s,t)\ = \ -{1\over 8\pi\beta^2 }\ \ ^0N(00;00) \nonumber \\
&&\rho^\pi_{\{S,V,T,A\}}(s,t)= 0 \nonumber
\end{eqnarray}
\begin{eqnarray}
\rho^\omega_S(s,t) &=&\ \, \ {1\over 16\alpha^2
\pi\beta}\cos\vartheta\nonumber \\ &&
\hspace{0.9cm} [\sqrt 2 (\beta-1)^2\ ^1N(01;21) + (2\beta^2+5\beta+2)\
^1N(21;21) -2(\beta-1)^2\ ^1N(01;01) ] \nonumber \\ \rho^\omega_V(s,t)&
=& \ - {1\over 16\alpha^2 \pi\beta}\ [\sqrt 2 (2\beta+1)\ ^1N(01;21) +
(\beta^2+2)\; ^1N(21;21) +2(\beta-1)\ ^1N(01;01)] \nonumber \\
\rho^\omega_T(s,t)&=& {1\over 128\,\alpha^2 \pi\beta^2\,m_N^2}\,t\,
\nonumber \\ &&
\hspace{0.9cm} [\sqrt 2 (\beta+2)\ ^1N(01;21) + (2\beta+1)\ ^1N(21;21)
-2(\beta-1)\ ^1N(01;01)] \nonumber \\ \rho^\omega_P(s,t)& =& {1\over
16\pi\beta^2} \cos\vartheta [\sqrt 2 (\beta+2)\ ^1N(01;21) + (2\beta+1)\
^1N(21;21) -2(\beta-1)\ ^1N(01;01)] \nonumber \\ \rho^\omega_{A}(s,t)&
=&\ \, \ 0 \nonumber
\end{eqnarray}
\begin{eqnarray}
&&\rho^{A_1}_P(s,t)\ = \ -{3\over 16\alpha^2\pi\beta^2}\ \ ^1N(11;11)
\nonumber \\ &&\rho^{A_1}_A(s,t)\ =\ -{3\over 16\alpha^2\pi}\ \
^1N(11;11)\nonumber \\ &&\rho^{A_1}_{\{S,V,T\}}(s,t) = 0 \nonumber \\
\nonumber \\ &&\rho^{H_1}_P(s,t)\ = \ -{3\over 8\pi\beta^2}\
\cos\vartheta \ \ ^1N(10;10) \nonumber \\
&&\rho^{H_1}_{\{S,V,T,A\}}(s,t) = 0 \\
\label{SpecFunc}
\nonumber
\end{eqnarray}

\noindent
with $\beta^2\equiv E^2(p)/m_N^2$ and $\alpha^2\equiv \beta^2-1$.

\subsection{Microscopic model for the $N\overline N\to\pi\rho$ process}

The determination of the spectral functions $\rho_i$ requires the
knowledge of the transition amplitude $t_{N\overline N\to\pi\rho}$
including $\pi\rho$ correlations.  In our dynamical model whose
structure is visualized in Fig.~\ref{figVPVGT} this quantity is obtained
from

\begin{equation}
t_{N\overline N\to\pi\rho}=v_{N\overline N\to\pi\rho}+ \,v_{N\overline
N\to\pi\rho} G_{\pi\rho}T_{\pi\rho\to\pi\rho}\ ,
\label{nnprlse}
\end{equation}

\noindent
where $v_{N\overline N\to\pi\rho}$ is the transition potential specified
later, $G_{\pi\rho}$ is chosen to be the Blankenbecler-Sugar
\cite{Blankenbecler66} propagator of the $\pi\rho$ system, and
$T_{\pi\rho\to\pi\rho}$ is the $\pi\rho\to\pi\rho$ amplitude essentially
taken from the dynamical model \cite{JanssenPRC94}. After partial wave
decomposition Eq.~(\ref{nnprlse}) reads more explicitly, in the helicity
state basis,

\begin{eqnarray}
&&t^J(k \lambda_\rho; p \lambda_N\lambda_{\overline N};\sqrt t) = v^J(k
\lambda_\rho; p \lambda_N\lambda_{\overline N};\sqrt t) \nonumber \\ &&
\ \ + \sum_{\lambda_\rho'}\int_0^\infty dk' k'^2
{\omega_\rho(k')+\omega_\pi(k')\over
(2\pi)^32\,\omega_\rho(k')\omega_\pi(k')} \ {v^J(k' \lambda_\rho'; p
\lambda_N\lambda_{\overline N};\sqrt t)\; T^J(k
\lambda_\rho;k'\lambda_\rho';\sqrt t) \over
t-(\omega_\rho(k')+\omega_\pi(k'))^2} \quad .
\end{eqnarray}

\subsubsection{The transition potential $v_{N\overline N\to\pi\rho}$.}

Our model for the transition potential $v_{N\overline N\to\pi\rho}$ is
based on nucleon and $\Delta$ exchange, together with an $\omega$ pole
term (Fig.~\ref{fignnprpot}).  In principle, further pole terms exist in
the channels considered in this work which are, however, not included in
the present calculations for the following reasons: In case of the pion,
its mass lies far below the $\pi\rho$ threshold so that such a diagram
has a negligible influence on the dispersion integral,
Eq.~(\ref{Disp}). In case of the $A_1$ and $H_1$ little is known about
their coupling strength to the nucleon. On the other hand, the (bare)
$\omega NN$ coupling constant can be fixed by adjusting the final result
to the empirical $NN$ repulsion (see below).

The starting point for the evaluation of the corresponding potential
expressions is the set of interaction Lagrangians

\begin{eqnarray}
 {\cal L}_{\pi NN}&=&\frac{f_{\pi NN}}{m_\pi}\overline\psi\gamma^5
\gamma^\mu\vec\tau\partial_\mu\vec\pi\psi \nonumber \\
\noalign{\vskip5pt} {\cal L}_{\rho NN}&=&g_{\rho NN}\ \overline
\psi\left[\gamma^\mu\vec \tau\psi\vec\rho_\mu
+\frac{\kappa}{4m_N}\sigma^{\mu\nu}\vec\tau(\partial_\mu\vec
\rho_\nu-\partial_\nu \vec\rho_\mu)\right]\psi \nonumber \\
\noalign{\vskip5pt} {\cal L}_{\pi \Delta N}&=&\frac{f_{\pi \Delta
N}}{m_\pi}\overline\psi \vec {\cal T}\psi_\mu\partial^\mu\vec \pi \ +\
h.c.  \nonumber \\ \noalign{\vskip5pt} {\cal L}_{\rho \Delta
N}&=&\frac{f_{\rho \Delta N}}{m_\rho}\overline\psi i \gamma^5\gamma^\mu
\vec {\cal T}\psi^\nu (\partial_\mu\vec \rho_\nu-\partial_\nu \vec
\rho_\mu)\ +\ h.c.  \nonumber \\ \nonumber \\ \noalign{\vskip5pt} {\cal
L}_{\omega NN}&=&g_{\omega NN}\ \overline
\psi\gamma^\mu\omega_\mu\psi\nonumber \\ \noalign{\vskip5pt} {\cal
L}_{\omega\pi\rho}&=&g_{\omega\pi\rho}\epsilon^{\mu\nu\sigma\tau}
\partial_\mu\omega_\nu\partial_\sigma\vec\rho_\tau\vec\pi
\label{Lagr}
\end{eqnarray}

We then obtain for the structure of the potential matrix elements

\medskip

\noindent
nucleon exchange:
\begin{equation}
v_s= i\;f\,F^2\,\frac{f_{NN\pi}g_{NN\rho}}{m_\pi}\ \frac{\bar v(q_2)
\left\{ \gamma^5 k\hspace{-5pt}/ _2\ [p\hspace{-5pt}/_x+m_N]\
(\epsilon\hspace{-5pt}/^*- \frac{\kappa}{4m_N}k\hspace{-5pt}/_1
\epsilon\hspace{-5pt}/^* +\frac{\kappa}{4m_N}\epsilon\hspace{-5pt}/^*
k\hspace{-5pt}/_1 ) \right\} u(q_1)}{p_x^2-m_N^2}
\end{equation}
\noindent
$\Delta$ exchange:
\begin{eqnarray}
&&v_s= -i\;f\,F^2\,\frac{f_{N\Delta\pi}g_{N\Delta\rho}}{m_\pi m_\rho}\
\bar v(q_2) \ (k_2)_\mu\ S^{\mu\nu}\ \gamma^5\gamma^\sigma \
[(k_1)_\nu\epsilon^*_\sigma-(k_1)_\sigma\epsilon^*_\nu]\ u(q_1)
\nonumber \\ &&\
S^{\mu\nu}=\frac{p\hspace{-5pt}/_x+m_\Delta}{p_x^2-m_\Delta^2}
\left\{-g^{\mu\nu}+\frac{1}{3}\gamma^\mu\gamma^\nu+\frac{2}{3m_\Delta^2}
p_x^\mu p_x^\nu-\frac{1}{3m_\Delta}(p_x^\mu p_x^\nu-p_x^\nu
p_x^\mu)\right\} \ \ \nonumber \\
\end{eqnarray}

\noindent
$\omega$ exchange:
\begin{equation}
v_t=-f\,F^2\,\frac{g^{(0)}_{\pi\rho\omega}g^{(0)}_{NN\omega}}
{m_\omega}\ \frac{\sqrt t}{p_x^2-(m^{(0)}_\omega)^2} \
\epsilon^{0\nu\sigma\tau}\,(k_1)_\sigma\,\epsilon^*_\tau\ \bar
v(q_2)\,\gamma_\nu\,u(q_1)
\end{equation}

\noindent
where $k_1$ ($k_2$) and $q_1$ ($q_2$) denote the four-momenta of the
$\rho$($\pi$) and nucleon (antinucleon), respectively. $p_x$ is the
momentum of the exchanged particle; for the $\omega$ exchange term, in
the cm system, $p_x=(\sqrt t, 0)$.  $\epsilon^*$ is the polarization
vector for the outgoing $\rho$ meson.  $F^2$ denotes the product of
vertex form factors, for which we used

\begin{eqnarray}
s{\rm-channel:}\ \ \ &&F^2=\left ( \frac{2\Lambda_{\pi
NX}^2-M^2_X}{2\Lambda_{\pi NX}^2-p_x^2} \right )^2 \left (
\frac{2\Lambda_{\rho NX}^2-M^2_X}{2\Lambda_{\pi NX}^2-p_x^2} \right )^2
\nonumber \\ t{\rm-channel:} \ \ \ \ &&F^2=\left( \frac{\Lambda_{\pi\rho
X}^2+m^2_X} {\Lambda_{\pi\rho X}^2+[\omega_\pi(k)+\omega_\rho(k)]^2}
\right )^2 \left( \frac{2\Lambda_{NN\, X}^2+m^2_X}{2\Lambda_{NN\,
X}^2+4E(p_x)^2} \right )^2 \ \ , \nonumber \\
\label{ffnnpr}
\end{eqnarray}

\noindent
where $X$ stands for the exchanged particle. $f$ denotes the isospin
factor; corresponding values are given in Table~\ref{tabparaI}. The
coupling constants are either experimentally known or fixed from our
former studies. An exception is the bare $\omega NN$ coupling
$g^{(0)}_{\omega NN}$ which, as mentioned before, will be fixed
later. Values for coupling constants and cutoff masses used are given in
Table~\ref{tabparaII}.  [The $s$-channel cutoff masses have been
adjusted to reproduce the overall strength of $N\overline N \to\pi\rho$
potential used in earlier studies~\cite{Mull91}.  This potential
produces good agreement with empirical information above the $N\overline
N$ threshold, but is based on a different off-shell behavior
(time-ordered perturbation theory rather than BbS).]  The zero-th
components of momenta are determined by the BbS reduction \cite{Aaron76}
to be $q_1^0=q_2^0=\sqrt t/2$, $k_1^0=\frac{1}{2}[\sqrt
t+\omega_\rho(k)-\omega_\pi(k)]$, and $k_2^0=\frac{1}{2}[\sqrt
t+\omega_\pi(k)-\omega_\rho(k)]$. The potential matrix elements are then
decomposed into LSJ partial waves in the standard way. Further
$u$-channel diagrams arising from $N$ and $\Delta$ exchange can be taken
into account by adding a factor of 2 in the (partial wave) $s$-channel
contributions.

\subsubsection{The amplitude $T_{\pi\rho\to\pi\rho}$.}

Our model for the correlation amplitude $T_{\pi\rho\to\pi\rho}$
\cite{JanssenPRC94} is generated from the three-dimensional BbS
\cite{Blankenbecler66} scattering equation

\begin{equation}
T({\bf k'},{\bf k};E)=V({\bf k'},{\bf k};E)+ \int\,d\,^3\,k''\, V({\bf
k'},{\bf k''};E) G({\bf k''};E)T({\bf k''},{\bf k};E) \ ,
\label{LSE}
\end{equation}
                                               \noindent ($\bf k$, $\bf
k'$, $\bf k''$ are corresponding cm relative momenta) with the potential
$V$ containing the diagrams shown in Fig.~\ref{figformI}. It contains,
besides non-pole pieces, pole terms with bare parameters (masses,
coupling constants) which are renormalized by the iteration in
Eq.~(\ref{LSE}) and in this way acquire their physical properties.
Basic interaction Lagrangians have been taken from the nonlinear
$\sigma$-model in the meson sector where the vector mesons are
introduced as gauge bosons of a hidden $SU(2)$ or $SU(3)$ symmetry. In
this way one obtains the coupling of the $\rho$ to the $\pi$ meson and
to itself, i.e.\ ${\cal L}_{\pi\pi\rho}$ and ${\cal L}_{\rho\rho\rho}$ ,
with a unified value for the coupling constants.  Note that we have left
out a corresponding pion pole term.  The reason is the very small pion
mass lying far below the $\pi\rho$ threshold, so that such a diagram
should have a negligible influence in the dispersion integral,
Eq.~(\ref{Disp}).

In addition to the model presented in Ref. \cite{JanssenPRC94} the
present calculations include the $H_1$ ($J^P=1^+$, $I^G=0^-$) channel;
therefore $V$ now contains an $H_1$-pole term. The corresponding
expression is analogous to the $A_1$-term, see \cite{JanssenPRC94}, with
the isospin factor $f=3\delta_{I,0}$, bare coupling constant
$(g^{(0)}_{H_1 \pi\rho})^2/4\pi=1.3$ and bare mass $m^{(0)}_{H_1}$ =
1100 MeV. As Fig.~\ref{figformII} shows, we obtain a reasonable
description of the $H_1$ mass distribution, although compared to the
$A_1$ case the model underestimates the empirical situation somewhat.
Still, the rough agreement should be sufficient to estimate the
relevance of the $H_1$ channel for the correlated $\pi\rho$ exchange
$NN$ interaction.

\subsection{$NN$ interaction arising from correlated $\pi\rho$ exchange}
\label{sect:NNpot}

In the last section we specified the dynamical model for the $N\overline
N \to\pi\rho$ amplitude, which yields the spectral functions
$\rho(s,t)$, Eq.~(\ref{SpecFunc}). The dispersion integral,
Eq.~(\ref{Disp}), then determines the invariant amplitudes
$c_j(t,s;t<0)$ and thus the scattering operator $\hat T$
(Eq.~(\ref{expans})). The various $NN$ potential contributions are then
obtained by sandwiching $\hat T$ between in- and outgoing spinors (cf.\
Eq.~(\ref{invampI})).

Such a calculation can be directly pursued for the $\pi$ and $A_1$
channel since these spectral functions do not depend on $\cos\vartheta$,
which is, in terms of the Mandelstam variables,

\begin{equation}
\cos\vartheta \ = \ {4m_N^2-t-2s \over t-4m_N^2}\ = \ {u-s \over
t-4m_N^2} \ \ .
\label{cos}
\end{equation}

When transforming into the $NN$ channel, corresponding values for $s>4
m_N^2$ have to be inserted, and, in principle, the $t$-dependence in
$\cos\vartheta$ should be integrated over in the dispersion
integral. However, it is then not guaranteed that the typical structure
of $s$-channel vector meson exchange is obtained. Namely, starting from
the conventional vector meson Lagrangian,

\begin{equation}
{\cal L}\ = \ g\, \bar \psi\gamma^\mu\psi\, V_\mu \;+\; f/4m_N\;\bar
\psi \sigma^{\mu\nu}\psi\,(\partial_\mu\,V_\nu
-\partial_\nu\,V_\mu)\quad,
\end{equation}

\noindent
one obtains for the scattering operator ($m_V$ is the mass of the vector
meson)

\begin{equation}
{\hat T} \, = \, {1\over t-m_V^2} \left\{g^2\,[-V]-gf\,[\,V+{u-s\over
4m_N^2}S+{t\over 8m_N^2}T +{u-s\over 4m_N^2}P\,]-f^2\,[\,{t\over 8m_N^2}
T + {u-s\over 4m_N^2}P\,]\right\} \ \ .
\label{Vectexneu}
\end{equation}

Obviously a characteristic factor $u-s \sim \cos\vartheta$ occurs in
front of the invariants $S$ and $P$ as well as a factor $t$ in front of
$T$. This structure is, for the case of $\rho$ exchange, of decisive
importance for a correct behavior of the $NN$ interaction. However,
since we have to apply approximations when evaluating the dispersion
integral (by introducing a cutoff $t_c = 4m_N^2$), this structure is not
automatically obtained when doing a straightforward
calculation. Therefore we decided to transform these factors directly
into the $NN$ channel and to apply the dispersion integral for the
remaining part of the spectral functions.  (In a more formal language,
new invariant operators have to be defined which include these factors.)
Trivially the results are then forced to have the structure of
$s$-channel vector meson exchange.

Another important modification of the above formulas remains to be
introduced. So far, by construction (see Fig.~\ref{figVPVGT}), our
results contain not only the correlated part we are interested in, but
also the uncorrelated contribution, cf.\ Fig.~\ref{figintI}. Therefore
the latter has to be removed. We do this by subtracting the Born term
part of the $^JN$ functions of Eq.~(\ref{LSJamps}), which leads to new
functions $^JN_{corr}$ given by

\begin{equation}
^JN_{corr}(00;00)=-C[(t^0_{0+})^\dagger t^0_{0+}-(v^0_{0+})^\dagger
v^0_{0+}]
\end{equation}

\noindent
for the pion and analogous extensions for the other channels. These new
functions $^JN_{corr}$ are actually used when evaluating the spectral
functions by means of Eq.~(\ref{SpecFunc}).

We still have to transform the isospin part of the $N\overline N\to
N\overline N$ amplitude into the $s$-channel. Resulting isospin factors
are provided by the isospin crossing matrix~\cite{Martin70}. In general
we have

\begin{eqnarray}
&&f^{I=0}_{NN} = {1\over 2}(f^{I=0}_{N\overline N} -3f^{I=1}_{N\overline
N}) \nonumber \\ &&f^{I=1}_{NN} = {1\over 2}(f^{I=0}_{N\overline N}
+f^{I=1}_{N\overline N})
\end{eqnarray}

\noindent
for the connection of isospin factors in both channels. The factors for
our $N\overline N\to N\overline N$ amplitude are already implicitly
taken into account by including corresponding factors in $v_{N\overline
N \to \pi\rho}$ and $T_{\pi\rho\to\pi\rho}$. For the (isospin zero)
$\omega$ and $H_1$ channels we thus have $f^I_{N\overline
N}=\delta_{I,0}$ whereas in $\pi$ and $A_1$ we have $f^I_{N\overline
N}=\delta_{I,1}$.  Therefore our final result for the correlated
$\pi\rho$ exchange $NN$ potential can be written as operator in isospin
space in the following way:

\begin{equation}
V_{\pi\rho,corr}=\frac{\kappa}{2}\sum_i \left [ \sum_{\alpha=\omega,H_1}
\int^{t_c}_{(m_\pi+m_\rho)^2}dt' \frac{\rho^\alpha_i(t')}{t'-t} C_i {\bf
1} +\sum_{\beta=\pi,A_1} \int^{t_c}_{(m_\pi+m_\rho)^2}dt'
\frac{\rho^\beta_i(t')}{t'-t} C_i {\bf \tau}_1\cdot {\bf \tau}_2 \right
] \quad,
\end{equation}

\noindent
where $C_i$, according to the foregoing discussion, are matrix elements
between nucleon helicity spinors of slightly modified invariants
$(u-s)S$, $(u-s)P$, $V$, $A$, $tT$. The factor
$\kappa=\frac{1}{(2\pi)^3} \frac{m_N^2}{\sqrt{E_1E_1'E_2E_2'}}$ arises
because $V_{\pi\rho,corr}$ is to be defined as part of the Bonn
potential whose $T$-matrix is defined by

\begin{equation}
S_{fi}=\delta_{fi}-i\,2\pi\,\delta^{(4)}(p_1'+p_2'-p_1-p_2) T_{fi} \ .
\end{equation}

In order to be used in a scattering equation the resulting potential
must be given off shell, as function of the in- and outgoing cm relative
momenta and total energy in the $s$ ($NN$) channel, i.e.\ $V =
V(p',p;E_{cm})$. On shell, for $p'=p=p_0$ with $E_{cm}^2 =
4(m_N^2+p_0^2)$, the relation of these variables to the Mandelstam
variables is unique and given by

\begin{eqnarray}
s&=&4E(p)^2 \nonumber \\ t&=&-2p^2(1-\cos\vartheta) \nonumber \\
u&=&-2p^2(1+\cos\vartheta) \ .
\label{on}
\end{eqnarray}

Half-off-shell, i.e.\ for $p'\ne p$, we take the plausible prescription
$t=-({\bf p}-{\bf p'})^2$ and $s=4E(p)E(p')$, which of course agrees
with Eq.~(\ref{on}) on shell. Dependence on the starting energy is
assumed to be of the same type as in time-ordered perturbation theory
applied in the Bonn potential. Here the propagator of an exchanged meson
reads

\begin{equation}
\frac{1}{\omega\, (E_{cm}-E(p)-E(p')-\omega)} \ .
\vspace{0.3cm}
\end{equation}

In order to obtain a natural generalization of this expression we first
define the `on-mass-shell energy of an exchanged $\pi\rho$ system',
$\Omega\equiv\sqrt{t'+({\bf p}-{\bf p'})^2}=\sqrt{t'-t}$ and replace the
energy denominator of the dispersion integral by the on-shell-equivalent
expression

\begin{equation}
\frac{1}{t'-t}\to\frac{1}{\Omega\, (E_{cm}-E(p)-E(p')-\Omega)} \ .
\vspace{0.3cm}
\end{equation}

Finally the resulting potentials have to fall off sufficiently rapidly
in order to be able to solve the scattering equation. For this reason we
introduce an additional form factor, $\left
(\frac{n\Lambda^2-m^2}{n\Lambda^2-t}\right )^n \to\left
(\frac{n\Lambda^2-t'}{n\Lambda^2-t} \right )^n$, with $\Lambda=$5 GeV,
$n$=5, into the dispersion integral. The large cutoff mass chosen
ensures that the results are not modified on shell.

\subsection{Determination of effective coupling constants and masses}
\label{sect:effcoup}

It is convenient to parametrize our correlated $\pi\rho$ exchange
results in terms of single, sharp-mass effective meson exchange, as done
for the analogous case of correlated $\pi\pi$ in the $NN$ \cite{Kim94}
and $\pi N$ \cite{Schuetz94} system.  For reasons to be discussed below,
this can only be done successfully for the $\pi$ and $\omega$ channel
contributions. For an effective $\pi'$ the scattering operator reads

\begin{equation}
\hat T= -g_{\pi'NN}^2\, \vec \tau_1\vec\tau_2\, \frac{1}
{t-m_{\pi'}^2}\, P
\end{equation}

For the $\omega'$, the expression has already been given in
Eq.~(\ref{Vectexneu}). By comparison of the coefficients belonging to
the invariants with the corresponding dispersion-theoretic terms one can
determine an effective coupling constant, which will in general be
$t$-dependent. For example we have for the pionic channel

\begin{equation}
-g_{\pi'NN}^2(t)\;\frac{1}{t-m_{\pi'}^2}\ = \ \frac{1}{2} \left [
{1\over\pi}\int\limits_{(m_\pi+m_\rho)^2}^\infty {\rho^\pi_P(t')\over
t'-t} dt'\right ]\ \ ,
\end{equation}

The effective mass $m_{\pi'}$ is chosen such that $g_{\pi' NN}$ becomes
essentially independent of $t$. It turns out that for the $H_1$ channel
such a mass cannot be found. For channels involving several invariants,
different coupling constants (and masses) are obtained which depend on
the specific invariant for which the comparison is made. Obviously such
a parametrization is successful if the resulting values only weakly (if
at all) depend on the invariant chosen. This is the case for the
$\omega$-channel but not for the $A_1$-channel.

\section{Results and discussion}
\label{sect:results}

Having introduced the necessary formalism for the evaluation of
correlated $\pi\rho$ exchange we now present the results and investigate
their consequences for the $NN$ interaction. We start with the results
obtained in the $t$-channel, i.e.\ for the $N\overline N \to\pi\rho$
amplitude and the spectral functions.  We then discuss the properties of
the resulting potential contribution, which is obtained from the
(dispersion-theoretic) transformation into the $s$-channel, and point
out its role within the Bonn meson exchange $NN$ interaction
\cite{Machleidt87}.

\subsection{The $\pi$ channel}

The pionic channel of correlated $\pi\rho$ exchange is of special
importance.  As mentioned in the introduction it is a natural candidate
to provide additional (short ranged) tensor force required to fit
empirical $NN$ data with models using a realistic, soft $\pi NN$ form
factor.

Fig.~\ref{figresI} shows the $N\overline N \to\pi\rho$ (on-shell)
potential $v_{N\overline N \to\pi\rho}$ and the corresponding amplitude
$t_{N\overline N \to\pi\rho}$ obtained from Eq.~(\ref{nnprlse}), which
contains the effect of $\pi\rho$ correlations. The purely imaginary
potential has a strong increase at the pseudophysical $\pi\rho$
threshold and a maximum near $t=50m_\pi^2$. $\pi\rho$ correlations
strengthen this maximum; in addition they generate a real part in the
amplitude. Note that these modifications act quadratically in the $NN\to
NN$ amplitude and therefore also in the $NN$ potential, so that the
final effect of $\pi\rho$ correlations is stronger than
Fig.~\ref{figresI} suggests.

Indeed, the spectral function $\rho_P^\pi(t)$ in Fig.~\ref{figresII}
demonstrates that the piece due to correlated $\pi\rho$ exchange has a
considerable strength, although it is smaller than the uncorrelated part
generated by the transition potential only. One has to keep in mind that
a good part of the latter contribution consists of iterative $\pi\rho$
box diagrams involving $NN$ intermediate states, which are not part of
$V_{NN}$ but are generated by the $NN$ scattering equation. Consequently
the role of uncorrelated processes in $V_{NN}$ is considerably smaller
than suggested from the figure.

The spectral function of the correlated part has a clear maximum at
about $t$=60 $m_\pi^2$, thus representing the mass distribution of a
broad, heavy effective particle with pionic quantum numbers. A rough
parametrization of this contribution by sharp-mass particle exchange
appears to require a mass of about 1 GeV, noticeably smaller than chosen
in Ref. \cite{Holinde90} for the phenomenological $\pi'$.

In the $s$-channel we first demonstrate the influence of the resulting
on-shell potentials in the $^3S_1-\,^3D_1$ and $^1S_0$ partial waves as
function of the nucleon lab energy. In Fig.~\ref{figresIII} the dotted
and dashed curves show the corresponding one-pion-exchange (OPE)
potentials; obviously there is a strong suppression of OPEP when the
$\pi NN$ cutoff mass is reduced from the value of 1.3 GeV used in the
full Bonn potential to 1 GeV.  Most importantly, the addition of the
potential due to correlated $\pi\rho$ exchange in the pionic channel to
the dotted curve restores the original tensor force strength; obviously
it is able to counterbalance the suppression induced by the smaller $\pi
NN$ cutoff mass. The $^3S_1-\,^3D_1$ partial wave is of special
importance in this connection since it exclusively contains a tensor
force component, which is decisive for a realistic description of
deuteron properties.

It should be added that the above result is in remarkable agreement with
a previous calculation of the $\pi NN$ form factor \cite{JanssenPRL94},
Fig.~\ref{figresIV}, which consistently used the same $\pi\rho$
$t$-matrix and $N\overline N \to\pi\rho$ transition potential, and
independently arrived at $\Lambda_{\pi NN}$ = 1 GeV.

As discussed in Sect.~\ref{sect:effcoup}, correlated $\pi\rho$ exchange
in the $\pi$ and $\omega$ channels can be parametrized by sharp-mass
one-boson-exchange (OBE) potentials, provided that the effective
coupling constants are allowed to become
$t$-dependent. Fig.~\ref{figresV} shows such coupling constants for the
exchange of a heavy $\pi'$ for various chosen masses of $\pi'$.
Obviously the coupling becomes $t$-independent for a mass of 1020 MeV,
quite near the maximum of the correlated spectral function in
Fig.~\ref{figresII}, and the resulting coupling constant is $g_{\pi'
NN}^2/4\pi\simeq 9$.  If we choose $m_{\pi'}$ =1200 MeV as in
\cite{Holinde90}, the resulting coupling constant is noticeably
$t$-dependent; its strength is much smaller than used in
\cite{Holinde90}. There are two reasons for this discrepancy: First the
authors of \cite{Holinde90} applied a $\pi' NN$ form factor, which
reduces the strength at $t$=0 (relevant for $NN$ scattering) by more
than a factor of 2. Second, the strength of the $\pi'$ was
phenomenologically chosen in \cite{Holinde90} to compensate for a much
softer $\pi NN$ form factor, with a cutoff mass of 800 MeV.  (Indeed a
much smaller value ($g_{\pi' NN}^2/4\pi$= 70) is sufficient to
compensate for a form factor with $\Lambda_{\pi NN}$ = 900
MeV~\cite{Haidenbauer94}.) Obviously correlated $\pi\rho$ exchange can
only partly explain the phenomenological $\pi'$; another possible
mechanism is correlated $\pi\sigma$ exchange \cite{Ueda92} (where
$\sigma$ stands for correlated $\pi\pi$ exchange in the S-wave channel).

In order to show that the compensation for a softer $\pi NN$ form factor
by correlated $\pi\rho$ exchange is valid not only for on-shell
potentials, but also for $NN$ amplitudes and observables, we extrapolate
the correlated $\pi\rho$ exchange off shell as described in
Sect.~\ref{sect:NNpot}, add this piece to the (full) Bonn potential
(with a reduced $\pi NN$ cutoff mass of 1 GeV) and solve the
relativistic Schroedinger equation relevant for the Bonn potential. Only
a slight readjustment of the coupling of the isospin-one scalar meson
($\delta$ in \cite{Machleidt87}) in the original Bonn potential is
required in order to obtain again a good description of $NN$ phase
shifts.

As Table~\ref{TABdeut} demonstrates convincingly the deuteron
observables also can be reproduced with a considerably softer $\pi NN$
form factor (characterized by $\Lambda_{\pi NN}$ = 1 GeV), provided that
correlated $\pi\rho$ exchange in the pionic channel is included.

\subsection{The $\omega$-channel}

Since the mass of the physical $\omega$-meson is only slightly below the
$\pi\rho$ threshold, genuine pole terms have been included in the
$\pi\rho$ amplitude \cite{JanssenPRC94} as well as in our model for the
transition potential $v_{N\overline N\to \pi\rho}$. As
Fig.~\ref{figresVI}(a) shows, the contribution of such pole terms leads
to a reduction of the (imaginary) transition potential above the
$\pi\rho$ threshold, whose amount depends on the value of the bare
coupling constant $g^{(0)}_{\omega NN}$.  (The reason for our choice
$g^{(0)}_{\omega NN}/4\pi$ = 4.40 will be discussed later.)
Fig.~\ref{figresVI}(b) shows the resulting amplitude $t_{N\overline N\to
\pi\rho}$.  Similarly to the pionic channel, $\pi\rho$ correlations
enhance the maximum of the amplitude near threshold.

The inclusion of the $\omega$-meson pole terms in our dynamical model
ensures that the imaginary part of the $N\overline N\to N\overline N$
amplitude, and therefore the resulting $NN$ potential, contains, besides
`true' correlated $\pi\rho$ exchange (Fig.~\ref{figresVII}(e)) generated
by the non-pole parts of the corresponding amplitudes, also genuine
$\omega$ exchange processes (Fig.~\ref{figresVII}(a)-(d)).
%(Note that the inclusion of diagrams of
%type (a)-(d) in the pionic channel would lead to double counting since
%they are already included in the conventional one-pion-exchange
%potential with the empirical $\pi NN$ coupling).
Corresponding propagators and vertex functions are dressed by $\pi\rho$
loop corrections.  For example, the $\omega$ propagator has the
following structure (for details we refer the reader to
\cite{JanssenPRC94}):

\begin{equation}
d=\frac{1}{t-(m^{(0)}_{\omega})^2-\Sigma(t)}, \quad {\rm with}\quad
\Sigma(t)\sim \int f^{(0)} G_{\pi\rho} f \quad,
\label{prop}
\end{equation}

\noindent
where $f^{(0)}$ and $f$ are bare and dressed $\omega\to\pi\rho$ vertex
functions, respectively, and $G_{\pi\rho}$ denotes the $\pi\rho$
propagator.

The bare parameters $g_{\omega\pi\rho}^{(0)}$ and $m^{(0)}_\omega$ have
been adjusted such that $d$ has a pole at $t=m_\omega^2$. The imaginary
part of the $N\overline N\to N\overline N$ amplitude, and therefore the
corresponding spectral functions, consist of a $\delta$-function at
$t=m_\omega^2$, which precisely corresponds to the exchange of a
(physical) $\omega$-meson with point-like $\omega NN$ coupling, i.e.\
without any form factor. However it is important to realize that $d$,
and therefore the diagrams in Fig.~\ref{figresVII}(a)-(d), provide a
further contribution, since $\pi\rho$ intermediate states make the
self-energy $\Sigma$ complex above the $\pi\rho$ threshold. Such
intermediate states likewise occur at the vertices in diagrams (b)-(d),
leading to additional contributions to the spectral functions.

Fig.~\ref{figresVIII} shows the resulting spectral functions. Note that
although we assumed the bare $\omega NN$ coupling to be of pure vector
type, small contributions to $\rho_S$, $\rho_T$, and $\rho_P$ occur,
which are generated by the $\pi\rho$ loops in
Fig.~\ref{figresVII}(b)-(d).  First we have the $\delta$-function piece
(dashed); above $\pi\rho$ threshold we have additional contributions
from diagrams \ref{figresVII}(a)-(d) (dash-dotted) which have opposite
sign to the $\delta$-function.  They act as vertex corrections which
suppress the point-like coupling of $\omega$ exchange and thus generate
a form factor effect. Finally there is a sizable non-pole contribution
(dotted curve); throughout it has opposite sign to the part generated by
the vertex corrections and roughly counterbalances their effect, a fact
found already in the pionic channel.

Again, after presenting the results in the $t$-channel, we now want to
look at the corresponding on-shell potentials in the $s$-channel.  Since
we deal with rather short-ranged contributions we show, as two
characteristic examples, the results for the $^1S_0$ and $^3P_1$ partial
waves (Fig.~\ref{figresIX}). Note that the bare $\omega NN$ coupling
($g^{(0)}_{\omega NN}$=4.40), which determines the size of diagrams
(a)-(c), has been chosen such that the total repulsion generated by all
diagrams of Fig.~\ref{figresVII} agrees (at low energies) with the
effective $\omega$ exchange in the Bonn potential needed
empirically. The dashed curves are generated by the $\delta$-functions
in Fig.~\ref{figresVIII}, with a predicted renormalized coupling
constant of $g^2_{\omega NN}$= 11.0. Apparently this contribution alone
provides almost the same repulsion as in the Bonn potential although the
coupling constant is about a factor of 2 smaller. The reason is that the
phenomenological form factor in the Bonn potential, with the monopole
cutoff mass (1.5 GeV) of only twice the $\omega$ mass, leads to a strong
reduction of the coupling constant in the physical region ($g^2_{\omega
NN}$($t$=0)=10.6).  The vertex corrections (generated by diagrams
(a)-(d) above $\pi\rho$ threshold) strongly reduce the repulsion in the
physical region, leading to the dash-dotted curve. This suppression is
essentially cancelled by the `true' correlated $\pi\rho$ exchange
(diagram \ref{figresVII} (e)), as already demonstrated for the spectral
functions. Obviously the latter contribution is remarkably strong; it
explains about 40\% of effective $\omega$ exchange.

The new reduced coupling constant (11.0) is still about a factor of 2
larger than provided by customary SU(3) estimates, which use
$g^2_{\omega NN}$= 9$g^2_{\rho NN}$. Thus with $g^2_{\rho NN}/4\pi$=0.55
as determined by Hoehler and Pietarinen \cite{Hoehler75} we have
$g^2_{\omega NN}/4\pi\simeq$5.  Note however that the above relation
between $\omega$ and $\rho$ coupling constants is based, apart from
ideal mixing, on the assumption of vanishing $\phi NN$ coupling. For
$g_{\phi NN}$ unequal to zero the above relation goes into

\begin{equation}
g_{NN\omega}=3g_{NN\rho}-\sqrt 2 g_{NN\phi} \ .
\end{equation}

Thus if we take $g_{\phi NN}$= -$g_{\rho NN}$ (which amounts to a rather
small deviation from zero) we have $g^2_{\omega NN}\approx 20 g^2_{\rho
NN}$, in rough agreement with our results. Such a value for the $\phi
NN$ coupling to the nucleon and the negative sign is well conceivable,
if the $\phi$ couples to the nucleon via the $K\overline K$ continuum
\cite{Mullpriv}.

As discussed before, for practical reasons it is convenient to
parametrize also the non-pole contribution of diagram \ref{figresVII}(e)
by an effective one-boson-exchange.  Results are shown in
Fig.~\ref{figresX}, for the dominant vector as well as the tensor
coupling. Obviously they can be reasonably represented by $g^2_{\omega'
NN}/4\pi$=8.5, $f_{\omega' NN}/g_{\omega' NN}$=0.4, and
$m_{\omega'}$=1120 MeV. Using the mass of the physical $\omega$ meson
$m_\omega=$782 MeV we find a ($t$-dependent) effective $\omega$ coupling
strength characterized by $g^2_{\omega NN} \simeq$ 4. It is interesting
to note that the suppression of the tensor coupling, in some sense
enforced in the pole terms by assuming the corresponding bare coupling
to be zero, also happens in the non-pole term.

\subsection{The $A_1$/$H_1$-channel}

After the discussion of the $\pi$ and $\omega$ channel we now want to
investigate the $A_1$ and $H_1$ channels together since their structure
is very similar. Compared to the $\pi$ and $\omega$ channels we have
important differences. First the $A_1$ as well as the $H_1$ mass lie
above the $\pi\rho$ threshold, and both particles decay with a very
large width into $\pi\rho$. Consequently their propagators now acquire a
pole in the complex plane. Second, in contrast to the $\pi$ and $\omega$
channels, the Bonn potential \cite{Machleidt87} does not contain
$A_1$/$H_1$ OBE contributions, which could be used to fix the bare $A_1
NN$ and $H_1 NN$ coupling constants, as was done in the $\omega$
channel. Since there is no other {\em a priori} information about these
couplings, we will in this first extrapolatory study, simply put them to
zero, i.e.\ take only diagrams of type \ref{figresVII}(d) and (e) into
account. If, in a later stage, those couplings turn out to be needed
(e.g.\ in order to obtain a quantitative fit to the $NN$ data), they
should be included.

There is a further structural difference which has an enormous impact on
the results and requires an extended discussion. In general, besides the
unitarity cut for $t>(m_\pi +m_\rho)^2$ treated in the dispersion
integral (Eq.~(\ref{Disp})), there exists a left hand cut for $t<t_0$ in
the $N\overline N \to \pi\rho$ amplitude generated by $s$-channel poles
due to nucleon and $\Delta$-isobar exchange (Fig.~\ref{fignnprpot}).
$t_0$ is fixed by the condition $s-m^2_{N/\Delta}$=0. There are two
solutions for each exchange; the largest value (generated by nucleon
exchange) is at $t_0\approx 42 m_\pi^2$, i.e.\ just below the $\pi\rho$
threshold.  Consequently this branch point will influence the resulting
potentials near threshold considerably.

In the $\pi$ and $\omega$ channels the corresponding potentials act in
P-waves and are thus proportional to the $\pi\rho$ on-shell momentum
$k_0$, with the effect that the corresponding transition potentials
start to increase first when approaching the threshold, but are then
suppressed by the $k_0$ factor. In this way one obtains the
characteristic structure of a maximum near threshold, which we have
observed in such channels.  The point now is that both the $A_1$ and
$H_1$ are $\pi\rho$ S-waves; therefore the transition potentials do not
contain the damping factor $k_0$ anymore.  Indeed, Figs.~\ref{figresXII}
and \ref{figresXIII} show the overwhelming effect of the left hand cut
near threshold, in both channels. It essentially remains when $\pi\rho$
correlations (which contain the $A_1$ resonance) are included, although
the influence of the $A_1$ is clearly seen.

Fig.~\ref{figresXIV} shows the resulting spectral functions
$\rho_P^{H_1}$, $\rho_A^{A_1}$, and $\rho_P^{A_1}$.  Again the strong
effect of the left hand cut near threshold is obvious.  Note also that
for $\rho_P$, $A_1$ and $H_1$ provide roughly similar contributions, but
with opposite sign.

In Fig.~\ref{figresXV} we present the resulting on-shell potentials in
some selected partial waves. For both S-states, the (attractive) $A_1$
contribution strongly dominates the result arising from the $H_1$
channel but is considerably smaller (as far as the modulus is concerned)
compared to the corresponding piece in the $\omega$ channel, cf.\
Fig.~\ref{figresX}.  For $^3P_1$ both contributions have opposite sign
and roughly cancel; the total result is negligible compared to the
$\omega$ channel.

In contrast to the $\pi$, $\omega$ channels discussed before, the above
results cannot be suitably parametrized in terms of sharp-mass
exchanges. In case of the $A_1$, no reasonable mass can be found which
works for both spectral functions; moreover all effective coupling
constants become strongly $t$-dependent. The basic reason is again the
dominance of the left hand cut, which destroys the conventional bump
structure of the spectral functions found in other channels of
correlated $\pi\pi$ and $\pi\rho$ exchange.

\section{Concluding remarks}
\label{sect:concl}

In this work we have determined the contribution to the $NN$ interaction
due to the exchange of a correlated $\pi\rho$ pair between two
nucleons. The correlations between $\pi$ and $\rho$ have been taken into
account using a realistic meson exchange model of the $\pi\rho$
interaction \cite{JanssenPRC94}.  In a first step we evaluated the
$t$-channel amplitude $N\overline N \to N\overline N$ including
$\pi\rho$ correlations. The transformation into the $s$-channel with the
help of dispersion-theoretic methods then yields the correlated
$\pi\rho$ exchange $NN$ potential. We have investigated four relevant
channels of the $\pi\rho$ system characterized by the quantum numbers of
the physical particles $\pi$, $\omega$, $A_1$, and $H_1$.

In the pionic channel correlated $\pi\rho$ exchange yields a
short-ranged contribution, which roughly corresponds to an exchange of a
heavy (effective) $\pi'$ with a mass of about 1 GeV. The additional
tensor force generated by this potential is sufficient to compensate for
a reduction of the $\pi NN$ cutoff mass $\Lambda_{\pi NN}$ from 1.3 GeV
to 1.0 GeV in the one-pion-exchange potential. For basic theoretical
reasons, such a reduction is highly welcome, since various models of
nucleon structure unanimously predict a rather soft $\pi NN$ form factor
characterized by $\Lambda_{\pi NN}\simeq 0.8\ {\rm GeV}$. Such a small
value might be reached if correlated $\pi\sigma$ exchange is included,
too, which is also missing in the Bonn potential. (As usual, $\sigma$
stands for a low mass correlated $\pi\pi$ pair in the $0^+$ channel.)
Thus it appears that in the Bonn potential~\cite{Machleidt87} the
one-pion-exchange potential together with a hard form factor is an
effective description of `true' one-pion exchange (with a soft form
factor) plus correlated $\pi\rho$ (and $\pi\sigma$) exchange in the
pionic channel.

In the $\omega$ channel, the exchange of a correlated $\pi\rho$ pair
also provides a sizable contribution to the $NN$ interaction. Since the
$\omega$ mass is near the $\pi\rho$ threshold we have included the
genuine $\omega$-meson explicitly and replaced the (effective) $\omega$
exchange in the Bonn potential by the resulting correlated $\pi\rho$
potential, which can be decomposed into a pole and a non-pole piece. The
former provides a microscopic model for `true' $\omega$ exchange leading
to a renormalized $\omega NN$ coupling constant, $g^2_{\omega NN}/4\pi$
= 11.0, which is about a factor of two smaller than the effective value
of 20 used in the Bonn potential.  Thus `true' correlated $\pi\rho$
exchange (Fig.~\ref{figintI}(e)) provides almost half of the empirical
repulsion needed in the $NN$ interaction; it can be parametrized by
sharp-mass $\omega'$ exchange with $g^2_{\omega'NN}\simeq 8.5$,
$f_{\omega' NN}/g_{\omega' NN}\simeq$ 0.4 and $m_{\omega '}\simeq$ 1120
MeV.

Our present result for the $\omega$ coupling constant ($g_{\omega
NN}^2\approx 20 g_{\rho NN}^2 $) is well compatible with $SU(3)$,
provided that there exists a small, negative $\phi NN$ coupling of
vector type, with the magnitude of the order of the
$\rho$-coupling. Such a $\phi NN$ coupling (especially the required
negative sign) occurs naturally if it is supposed to arise via the
$K\overline K$ continuum. Although corresponding $\phi$ exchange in the
$NN$ interaction provides only a small contribution to the repulsion, it
makes the above relation between $\omega$ and $\rho$ couplings agree
with $SU(3)$ predictions. Consequently there appears to be little room
for explicit quark-gluon effects in being responsible for the
short-range $NN$ repulsion.

Additional contributions arise in the $A_1$ and $H_1$ channels. They are
sizable individually, mainly due to left-hand cut effects arising from
nucleon and $\Delta$ exchange in a $\pi\rho$ S-wave. However, in some
partial waves strong cancellations occur between the $A_1$ and $H_1$
contributions. Further contributions are, in principle, generated by
direct coupling of the $A_1$/$H_1$ to the nucleon. The size of such
couplings is, however, not known; these terms are therefore omitted in
the present work. It remains to be seen whether a precise fit of the
$NN$ observables requires such terms and thereby establishes their
existence.

\begin{appendix}
\section{Determination of spectral functions}

In this appendix we derive Eq.~(\ref{SpecFunc}) of the main text, which
provides the connection between the spectral functions needed in
Eq.~(\ref{Disp}) and the imaginary part of the $N\overline N\to
N\overline N$ amplitude in the LSJ basis ($^JN$) obtained from the
unitarity relation (Eq.~(\ref{unit})). For the latter it was necessary
to work in the LSJ basis in order to identify the allowed $N\overline N
\to \pi\rho$ transitions. In order to establish the connection to the
spectral functions however, matrix elements of the invariants $\hat C_j$
are required, which is most suitably done in the helicity state basis.
The imaginary part of the $N\overline N\to N \overline N$ amplitude is
now defined by (cf.\ Eq.~(\ref{UniRel}) for the analogous definition in
LSJ basis) $^JN( \lambda_N' \lambda_{\overline
N}';\lambda_N\lambda_{\overline N})\equiv {\rm Im}\,T^J (p' \lambda_N'
\lambda_{\overline N}'; p \lambda_N\lambda_{\overline N};\sqrt t)$ and
the relation between LSJ and helicity basis amplitudes is given by the
standard expressions (cf.\ \cite{Machleidt87})

\begin{eqnarray}
^J N_1&=&\ \frac{1}{2}\ ^JN(J0;J0)+a^2\ ^JN(J-1 1;J-11)-ab\
^JN(J+11;J-11) \nonumber \\ && \ \ \ \ -ab\ ^JN(J-11;J+11)+b^2 \
^JN(J+11;J+11) \nonumber \\ ^JN_2&=&-\frac{1}{2}\ ^JN(J0;J0) +a^2\
^JN(J-1 1;J-11)-ab\ ^JN(J+11;J-11) \nonumber \\ && \ \ \ \ -ab\
^JN(J-11;J+11)+b^2 \ ^JN(J+11;J+11) \nonumber \\ ^J N_3&=&\ \
\frac{1}{2}\ {^JN(J1;J1)}+b^2\ ^JN(J-1 1;J-11)+ab\ ^JN(J+11;J-11)
\nonumber \\ && \ \ \ \ +ab\ ^JN(J-11;J+11)+a^2 \ ^JN(J+11;J+11)
\nonumber \\ ^JN_4&=&-\frac{1}{2}\ {^JN(J1;J1)}+b^2\ ^JN(J-1 1;J-11)+ab\
^JN(J+11;J-11) \nonumber \\ && \ \ \ \ +ab\ ^JN(J-11;J+11)+a^2 \
^JN(J+11;J+11) \nonumber \\ ^J N_5&=&\ \ ab \ ^JN(J-1 1;J-11)-b^2\
^JN(J+11;J-11) +a^2\ ^JN(J-11;J+11) \nonumber \\ && \ \ \ \ -ab \
^JN(J+11;J+11) \nonumber \\ ^JN_6 &=&\ ^JN_5 \ \ \ \ (\rm on-shell)
\label{bla}
\end{eqnarray}
where we used the short-hand notation for the $^JN$ amplitudes defined
in Table~\ref{bull} and
\begin{equation}
a=\sqrt{{J\over 2(2J+1)}} \ \ \ \ \ \ \ b=\sqrt{{J+1 \over 2(2J+1)}} \ \
{}.
\end{equation}

\noindent
The various channel contributions to ${\rm Im} T({\bf
p'}\lambda_N'\lambda_{\overline N}'; {\bf p} \lambda_N\lambda_{\overline
N};\sqrt t)$ are then given by

\begin{eqnarray}
\pi:&& N_i\equiv N( \lambda_N' \lambda_{\overline
N}';\lambda_N\lambda_{\overline N})\equiv
\frac{1}{4\pi}d^0_{\lambda\lambda'}(\cos\vartheta)\ ^0N( \lambda_N'
\lambda_{\overline N}';\lambda_N\lambda_{\overline N})
\equiv\frac{1}{4\pi}d^0_{\lambda\lambda'}(\cos\vartheta)\ ^0N_i
\nonumber \\ \omega,A_1,H_1:&& N_i\equiv N( \lambda_N'
\lambda_{\overline N}';\lambda_N\lambda_{\overline N})\equiv
\frac{3}{4\pi}d^1_{\lambda\lambda'}(\cos\vartheta)\ ^1 N(\lambda_N'
\lambda_{\overline N}';\lambda_N\lambda_{\overline N}) \equiv
\frac{3}{4\pi}d^1_{\lambda\lambda'}(\cos\vartheta)\ ^1N_i \nonumber \\
\end{eqnarray}

We now define a vector $ {\bf N}^\alpha=\left (
\begin{array}{ccccc}
N_1^\alpha & N_2 ^\alpha& N_3^\alpha & N_4^\alpha & N_5^\alpha \\
\end{array}
\right ) $, $\alpha=\pi,\omega,A_1, H_1$,

\noindent
and use Eqs.~(\ref{LSJamps}) and (\ref{bla}) to express its components
in terms of the LSJ amplitudes $^JN(L'S';LS)$ for each of the
contributing channels. We obtain

\begin{equation}
{\bf N}^\pi=\left (
\begin{array}{c}
\frac{1}{8\pi}\ ^0N(00;00) \\-\frac{1}{8\pi}\ ^0N(00;00) \\ 0 \\ 0 \\ 0
\end{array}
\right ) \quad,
\end{equation}
\begin{equation}
{\bf N}^{A_1}=\left (
\begin{array}{c}
0 \\ 0 \\ \frac{3(1+\cos\vartheta)}{16\pi}\ ^1N(11;11) \\
\frac{3(-1+\cos\vartheta)}{16\pi}\ ^1N(11;11) \\ 0
\end{array}
\right ) \quad ,
\end{equation}
\begin{equation}
{\bf N}^{H_1}=\left (
\begin{array}{c}
\frac{3\cos\vartheta}{8\pi}\ ^1N(00;00) \\ -\frac{3\cos\vartheta}{8\pi}\
^1N(00;00) \\ 0 \\ 0 \\ 0
\end{array}
\right ) ,
\end{equation}
\begin{equation}
{\bf N}^\omega=\left (
\begin{array}{c}
\frac{3}{4\pi}\cos\vartheta [\frac{1}{6}\ ^1N(01;01)
-\sqrt{\frac{2}{9}}\ ^1N(01;21)+\frac{1}{3}\ ^1N(21;21)] \\
\frac{3}{4\pi}[\frac{1}{6}\ ^1N(01;01) -\sqrt{\frac{2}{9}}\
^1N(01;21)+\frac{1}{3}\ ^1N(21;21)] \\
\frac{1}{16\pi}(1+\cos\vartheta)[2\ ^1N(01;01) +2\sqrt{2}\ ^1N(01;21)+\
^1N(21;21)] \\ -\frac{1}{16\pi}(-1+\cos\vartheta)[2\ ^1N(01;01)
+2\sqrt{2}\ ^1N(01;21)+\ ^1N(21;21)] \\ \frac{1}{16\pi}\sin\vartheta[-2\
^1N(01;01) +\sqrt{2}\ ^1N(01;21)+2\ ^1N(21;21)]
\end{array}
\right )\ \ .
\end{equation}

The decomposition of the imaginary part of Eq.~(\ref{expans}) can now be
written for the helicity state matrix-elements in matrix notation

\begin{equation}
{\bf N}^{\pi,\omega,A_1,H_1} =\left (
\begin{array}{ccccc}
S_1 & V_1 & T_1 & P_1 & A_1 \\ S_2 & V_2 & T_2 & P_2 & A_2 \\ S_3 & V_3
& T_3 & P_3 & A_3 \\ S_4 & V_4 & T_4 & P_4 & A_4 \\ S_5 & V_5 & T_5 &
P_5 & A_5
\end{array}
\right ){\bf R}\equiv {\bf M} {\bf R}
\end{equation}

\noindent
where ${\bf R}=(\rho_S \ \rho_V \ \rho_T \ \rho_P \ \rho_A)$.  $S_i$,
$V_i$, $T_i$, $P_i$, and $A_i$ are helicity state matrix-elements of the
Fermi invariants using the same indexing as for the $N$ amplitudes and
we find

\begin{equation}
{\bf M} =\left (
\begin{array}{ccccc}
\beta^2-1 & -\cos\vartheta& -2\cos\vartheta & -\beta^2 & 1 \\ \beta^2-1
& -\cos\vartheta& 2\cos\vartheta(-2\beta^2+1) & \beta^2 & -1 \\ 0 &
-\beta^2(1+\cos\vartheta) & -2(1+\cos\vartheta) & 0 &
-(\beta^2-1)(1+\cos\vartheta) \\ 0 & -\beta^2(1-\cos\vartheta) &
-2(1-\cos\vartheta) & 0 & (\beta^2-1)(1-\cos\vartheta) \\ 0 &
\beta\sin\vartheta & 2\beta\sin\vartheta & 0 & 0
\end{array}
\right )
\end{equation}

The spectral functions are then simply obtained by calculating ${\bf R
}={\bf M}^{-1}{\bf N}^{\pi,\omega,A_1,H_1}$ yielding the result of
Eq.~(\ref{SpecFunc}).
\end{appendix}

%%%\bibliography{refs}
%%%\bibliographystyle{prsty}

\begin{figure}
\caption{$\pi\rho$ contributions to the $NN$ potential. (a) is generated
by the scattering equation whereas (b) and (c) are explicitly contained
in the Bonn potential. (d) shows correlated $\pi\rho$ exchange treated
in the present work.}
\label{figintI}
\end{figure}

\begin{figure}
\caption{Diagram visualizing the $NN$ scattering process, in the $s$
($NN\to NN$) and $t$ ($N\overline N\to N\overline N$) channels.}
\label{figone}
\end{figure}

\begin{figure}
\caption{Diagram visualizing the $\pi\rho$ exchange contribution to the
$NN$ interaction.}
\label{figtwo}
\end{figure}

\begin{figure}
\caption{The transition amplitude $t_{N\overline N\to \pi\rho}$.}
\label{figVPVGT}
\end{figure}

\begin{figure}
\caption{The transition potential $v_{N\overline N\to \pi\rho}$.}
\label{fignnprpot}
\end{figure}

\begin{figure}
 \caption{Contributions to the $\pi\rho$ potential.}
\label{figformI}
\end{figure}

\begin{figure}
\caption{Results of our $\pi\rho$ interaction model for the mass
distribution in the $H_1$ and $A_1$ channels compared with experiment
\protect\cite{Dankowych81}.}
\label{figformII}
\end{figure}

\begin{figure}
\caption{The (imaginary) transition potential $v_{N\overline N
\to\pi\rho}$ (dashed curve) in the pion channel together with the
corresponding amplitude $t_{N\overline N \to\pi\rho}$. The dotted line
shows the real part of $t$, the solid line the imaginary part.}
\label{figresI}
\end{figure}

\begin{figure}
\caption{The spectral function in the pionic channel
$\rho^\pi_P(t)$. The dotted line shows the uncorrelated part whereas the
solid line represents the correlated contribution.}
\label{figresII}
\end{figure}

\begin{figure}
\caption{On-shell $NN$ potential $V_{NN}$ as function of the nucleon lab
         energy in the $^1S_0$ state (a) and $^3D_1\,-\,^3S_1$
         transition (b). The dotted line denotes the one-pion-exchange
         potential as used in the Bonn potential ($g^2_{\pi NN}/4\pi =
         14.4,\ \Lambda_{\pi NN}= 1.3\ GeV$).  For the dashed line,
         $\Lambda_{\pi NN}=1.0\ GeV$. The solid line results if
         correlated $\pi\rho$ exchange in the pionic channel is added to
         the dashed line.}
\label{figresIII}
\end{figure}

\begin{figure}
\caption{Diagrams contributing to the $\pi NN$ form factor.}
\label{figresIV}
\end{figure}

\begin{figure}
\caption{The $t$-dependent effective $\pi' NN$ coupling constant for
various values of $m_{\pi'}$.}
\label{figresV}
\end{figure}

\begin{figure}
\caption{The (imaginary) transition potential $v_{N\overline N
\to\pi\rho}$ (dashed curve) in the $\omega$ channel (a) together with
the corresponding amplitude $t_{N\overline N \to\pi\rho}$ (b). The
dotted line shows the real part of $t$, the solid line the imaginary
part.}
\label{figresVI}
\end{figure}

\begin{figure}
\caption{Contributions to the $NN$ potential in the $\omega$ channel.}
\label{figresVII}
\end{figure}

\begin{figure}
\caption{The spectral functions in the $\omega$ channel. As explained in
the text the dashed line contains only the $\delta$-function
contribution of diagrams \protect\ref{figresVII} (a)-(d) whereas the
dash-dotted line shows the vertex and propagator corrections to the
corresponding diagrams.  The dotted line contains only the non-pole part
(\protect\ref{figresVII} (e)) and the solid line shows the full result.}
\label{figresVIII}
\end{figure}

\begin{figure}
\caption{On-shell $NN$ potential $V_{NN}$ as function of the nucleon lab
         energy in the $^1S_0$ and $^3P_1$ state. The long-dashed line
         shows the result of single $\omega$ exchange as contained in
         the Bonn potential ($g^2_{\omega NN}/4\pi$=20;$\Lambda_{\omega
         NN}$=1500 MeV). The dashed line contains only the
         $\delta$-function contribution of the diagrams of
         Figs.~\protect\ref{figresVII}(a)-(d) (cf.\ text) from which the
         dash-dotted line is obtained by adding the vertex and
         propagator corrections contained in the corresponding diagrams.
         The full result is shown by the solid line and contains in
         addition the non-pole part (Fig.~\protect\ref{figresVII}(e)).}
\label{figresIX}
\end{figure}

\begin{figure}
\caption{The $t$-dependent effective $\omega' NN$ vector coupling
constant $g^2_{\omega' NN}/4\pi$ and the ratio $f_{\omega'
NN}/g_{\omega' NN}$ for $m_{\omega'}$=1120 MeV (solid). The dashed line
shows $g^2_{\omega' NN}/4\pi$ when the physical $\omega$ mass (782.6
MeV) is used.}
\label{figresX}
\end{figure}

\begin{figure}
\caption{The (real) transition potential $v_{N\overline N \to\pi\rho}$
(dashed curve) in the $A_1$ channel together with the corresponding
amplitude $t_{N\overline N \to\pi\rho}$. The dotted line shows the real
part of $t$, the solid line the imaginary part.}
\label{figresXII}
\end{figure}

\begin{figure}
\caption{Same as Fig.~\protect\ref{figresXII} but for the $H_1$
channel.}
\label{figresXIII}
\end{figure}

\begin{figure}
\caption{The spectral functions in the $A_1$ and $H_1$ channel
((a):$\rho^{H_1}_P$; (b):$\rho^{A_1}_P$;(c):$\rho^{A_1}_A$). The dashed
line shows the contribution of the pole part whereas the dotted line
contains the non-pole part. The full result is given by the solid line.}
\label{figresXIV}
\end{figure}

\begin{figure}
\caption{On-shell $NN$ potential $V_{NN}$ as function of the nucleon lab
         energy in the $^1S_0$, $^3S_1$, and $^3P_1$ states. The
         dash-dotted (dotted) line shows correlated $\pi\rho$ exchange
         in the $A_1$ ($H_1$) channel.  }
\label{figresXV}
\end{figure}

\begin{table}[b]
\vskip0.4cm
\caption{
\label{Tabone}{Quantum numbers and possible
transitions from the $N\overline N$ to the $\pi\rho$ system.}}
\begin{tabular}{ccccccccc}
 & $J^P$ & $(I^{G})$ & $L_{\pi\rho}$ & $S_{\pi\rho}$ & $L_{N\overline
 N}$ & $S_{N\overline N}$ & $L_{N\overline N}S_{N\overline N}\to
 L_{\pi\rho} S_{\pi\rho}$ & notation \\ \noalign{\vskip3pt} \hline
 \noalign{\vskip3pt} $\pi$ & $0^-$ & ($1^-$) & 1 & 1 & 0 & 0 & J \ 0
 $\to$J+1\ 1 & $t^J_{0+}$ \\ $\omega$ & $1^-$ & ($0^-$) & 1 & 1 & 0 & 1
 & J$-$1 \ 1 $\to$J\ 1 & $t^J_{-1}$ \\ & & & 1 & 1 & 2 & 1 & J+1 \ 1
 $\to$J\ 1 & $t^J_{+1}$ \\ $A_1$ & $1^+$ & ($1^-$) & 0 & 1 & 1 & 1 & J \
 1 $\to$J$-$1\ 1 & $t^J_{1-}$ \\ & & & 2 & 1 & 1 & 1 & J \ 1 $\to$J+1\ 1
 & $t^J_{1+}$ \\ $H_1$ & $1^+$ & ($0^-$) & 0 & 1 & 1 & 0 & J \ 0
 $\to$J$-$1\ 1 & $t^J_{0-}$ \\ & & & 2 & 1 & 1 & 0 & J \ 0 $\to$J+1\ 1 &
 $t^J_{0+}$ \\
\end{tabular}
\end{table}

\begin{table}[h]
\caption{
\label{tabparaI}{Isospin factors for the $N\overline N\to\pi\rho$ transition
potential.}}
\begin{tabular}{cccc}
Exchange & Type of & \multicolumn{2}{c}{$f$} \\ particle & diagram &
$I$=0 & $I$=1 \\ \noalign{\vskip3pt} \hline \noalign{\vskip3pt} $N$ &
$s$ & -$\sqrt 6$ & -2 \\ $\Delta$ & $s$ & -$\sqrt{\frac{8}{3}}$ &
$\frac{2}{3}$ \\ $\omega$ & $t$ & -$\sqrt6$ & 0
\end{tabular}
\end{table}

\begin{table}[h]
\caption{
\label{tabparaII}{Coupling constants and cutoff masses
for the $N\overline N\to\pi\rho$ transition potential.}}
\begin{tabular}{cccc}
Vertex & $g^2/4\pi$ & $\kappa$ & $\Lambda$ [MeV] \\ \noalign{\vskip3pt}
 \hline \noalign{\vskip3pt} $NN \pi$ & 14.4 & - & 1970 \\ $NN \rho$ &
 0.84 & 6.1 & 1970 \\ $N\Delta\pi$ & 0.36 & - & 1150 \\ $N\Delta\rho$ &
 20.45 & - & 1150 \\ $NN \omega$ & 19.42 & - & 1414
\end{tabular}
\end{table}

\begin{table}
\caption{Deuteron properties predicted by the model including a soft
$\pi NN$ form factor ($\Lambda_{\pi NN}=1 {\rm GeV}$) and correlated
$\pi\rho$ exchange in comparison to the full Bonn potential and
experiment.}
\label{TABdeut}
\begin{tabular}{c|ccc}
 & present & full Bonn & experimental \\ & model & potential &
 data\tablenote{References are given in \protect\cite{Machleidt87}.}  \\
 \hline Binding energy $\epsilon_d$ [MeV] & 2.2246 & 2.2247 & 2.2245754
 \\
%\ (Ref. \protect\cite{LeunAlder})\\
% & & & $\pm$ 0.000009 \\
D state probability $P_d$ [\%] & 4.37 & 4.25 & - \\ Quadrupole moment
$Q_d$ [$fm^2$] & 0.2781 & 0.2807 & 0.2859$\pm$0.0003 \\
% (Ref. \protect\cite{BishCheung})\\
%Asymptotic S state $A_S$ [$fm^{-1/2}$] & 0.9027 & 0.9046
%& 0.8846$\pm$0.0016 \ (Ref. \protect\cite{ErRo})\\
Asymptotic D/S ratio & 0.0265 & 0.0267 & 0.0271$\pm$0.0008 \\
% (Ref. \prect\cite{RodnKnut})\\
\end{tabular}
\end{table}

\begin{table}[hb]
\caption{
\label{bull}{$NN$ matrix elements in helicity basis.}}
\begin{tabular}{c}
$<\lambda_N'\lambda_{\overline N}'|^JN| \lambda_N\lambda_{\overline N}>$
($\lambda=\pm\frac{1}{2})$\\ \noalign{\vskip3pt} \hline
\noalign{\vskip3pt} $^JN_1\equiv <+ + |^JN|+ + >$ \\ $^JN_2\equiv <+ +
|^JN|- - >$ \\ $^JN_3\equiv <+ - |^JN|+ - >$ \\ $^JN_4\equiv <+ - |^JN|-
+ >$ \\ $^JN_5\equiv <+ + |^JN|+ - >$ \\ $^JN_6\equiv <+ - |^JN|+ + >$
\\
\end{tabular}
\end{table}

\end{document}